\documentclass{Interspeech}

\interspeechcameraready

\title{Lessons Learnt: Revisit Key Training Strategies for Effective Speech Emotion Recognition in the Wild}

\author[affiliation={1}]{Jing-Tong}{Tzeng}
\author[affiliation={1, 2}]{Bo-Hao}{Su}
\author[affiliation={1}]{Ya-Tse}{Wu}
\author[affiliation={1}]{Hsing-Hang}{Chou}
\author[affiliation={1}]{Chi-Chun}{Lee}

\affiliation{Department of Electrical Engineering}{National Tsing Hua University}{Taiwan}
\affiliation{Language Technologies Institute}{Carnegie Mellon University}{USA}
\email{roger37890426@gmail.com, borrissu@gapp.nthu.edu.tw, crowpeter@gapp.nthu.edu.tw, stargazer@gapp.nthu.edu.tw, cclee@ee.nthu.edu.tw}
\keywords{speech emotion recognition, human-computer interaction, computational paralinguistics, valence recognition}

\usepackage{comment}
\usepackage[table,xcdraw]{xcolor}
\usepackage{pifont}
\usepackage{varwidth}
\usepackage{tcolorbox}
\newcommand{\cmark}{\ding{51}}
\newcommand{\xmark}{\ding{55}}

\begin{document}

\maketitle

\begin{abstract}

    In this study, we revisit key training strategies in machine learning often overlooked in favor of deeper architectures. Specifically, we explore balancing strategies, activation functions, and fine-tuning techniques to enhance \emph{speech emotion recognition} (SER) in naturalistic conditions. Our findings show that simple modifications improve generalization with minimal architectural changes. Our multi-modal fusion model, integrating these optimizations, achieves a valence CCC of 0.6953, the best valence score in Task 2: Emotional Attribute Regression. Notably, fine-tuning RoBERTa and WavLM separately in a single-modality setting, followed by feature fusion without training the backbone extractor, yields the highest valence performance. Additionally, focal loss and activation functions significantly enhance performance without increasing complexity. These results suggest that refining core components, rather than deepening models, leads to more robust SER in-the-wild.
\end{abstract}

\section{Introduction}
With the advancement of affective computing, \textit{Speech Emotion Recognition} (SER) has gained significant attention due to its potential in \textit{human-computer interaction} (HCI) \cite{picard2000affective, alnuaim2022human, ramakrishnan2013speech}. As SER continues to be integrated into real-world applications, including call center analytics \cite{yurtay2024emotion, bojanic2020call}, healthcare monitoring \cite{kumar2024comprehensive, james2021empathetic}, and digital assistants \cite{chatterjee2021real}, ensuring its robustness in unconstrained environments is crucial. To improve SER’s applicability in these settings, researchers have increasingly focused on "in-the-wild" dataset collection, where speech emotion data is captured under real-world conditions rather than controlled laboratory environments \cite{lotfian2017building, 9414542, Naini_2025}. However, these datasets introduce significant modeling challenges, as spontaneous emotional expressions vary across speakers, acoustic conditions, and contextual factors. A promising approach to addressing these challenges is \textit{self-supervised learning} (SSL), which enables models to extract robust speech representations from large-scale unlabeled data, thereby enhancing generalization across diverse conditions \cite{9053569, morais2022speech, kakouros2023speech}.

Despite these advancements, audio-only SSL features exhibit inherent limitations, particularly in capturing the valence dimension, which is often conveyed more effectively through textual and contextual information rather than acoustic cues alone \cite{sridhar2022unsupervised, wagner2023dawn, 10832143}. This limitation has driven recent research toward multi-modal approaches, where SSL-derived textual features are integrated with speech signals to develop dual-modality emotion recognition systems \cite{deoliveira23_interspeech, macary2021use, chu2022self}. However, key architectural aspects remain underexplored, including the impact of activation function choices in the fusion stage and whether fine-tuning single-modality SSL models contributes to improving overall system performance. Addressing these gaps is crucial for optimizing multi-modal SER frameworks and ensuring their robustness in real-world applications.


In this work, we revisit key training strategies for SER systems, with a particular focus on the fusion stage of dual-modality emotion recognition. Our contributions are as follows:
\begin{enumerate}
\item The effect of fine-tuning SSL features on individual modalities in multi-modality settings.
\item A comparison of activation functions.
\item An investigation into the impact of textual features between SSL and \textit{large language models} (LLMs).
\end{enumerate}

To this end, we fine-tune WavLM and RoBERTa on their respective modalities and optimize fusion strategies, both guided by insights gained from our analysis. As a result, our model achieves the best \textit{concordance correlation coefficient} (CCC) of 0.6953 on valence in the Speech Emotion Recognition in Naturalistic Conditions Challenge.

\begin{figure*}[t]
  \centering
  \includegraphics[width=\linewidth]{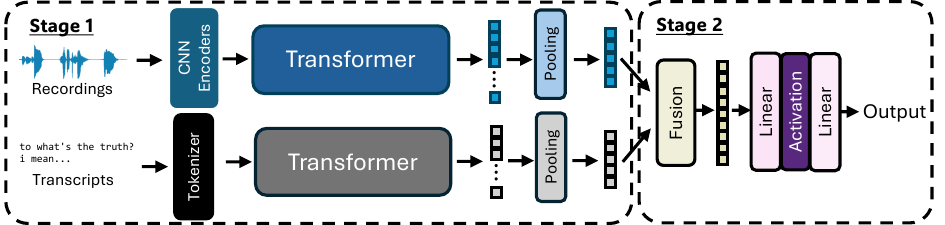}
  
  \caption{Our framework of the dual-modality (acoustic and textual) SER system}
  \label{fig:flowchart}
\end{figure*}

\section{Related Works}

It is widely recognized that arousal and dominance are more speech-dependent, whereas valence relies more heavily on textual information \cite{wagner2023dawn, 10832143}. As a result, many studies have explored dual-modality fusion between acoustic and textual features to improve SER.
For instance, Macary \textit{et al.} demonstrate that combining self-supervised models such as wav2vec \cite{schneider19_interspeech} and camemBERT \cite{martin2020camembert} enhances continuous speech emotion recognition \cite{macary2021use}. Similarly, Chu \textit{et al.} propose a deeply fused audio-text transformer with stage-wise cross-modal pretraining, improving both SER and sentiment analysis \cite{chu2022self}.

A key observation across these studies is the critical role of SSL features; however, approaches differ on whether foundation models require fine-tuning for emotion recognition and whether fine-tuning is necessary during fusion. Given the "in-the-wild" nature of our dataset, this work investigates modality-specific fine-tuning and examines the impact of activation functions in the fusion layer, contributing to a deeper understanding of optimal fusion strategies for SER.










\begin{figure*}[t]
  \centering
  \includegraphics[width=\linewidth]{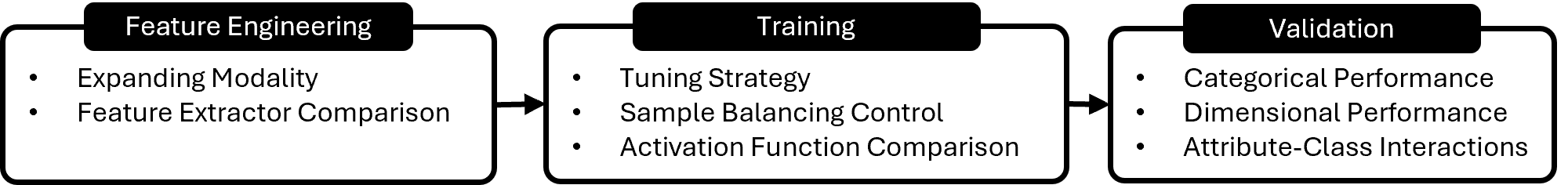}
  \caption{Common Machine Learning Pipeline}

  \label{fig:Common}
  \vspace{-3mm}
\end{figure*}

\begin{table*}[ht]
\centering
\setlength{\tabcolsep}{4pt}
\begin{tabular}{l|c|c|c|c|ccc|ccc|c}
\toprule
                       & \multicolumn{2}{c|}{Speech modality fine-tune} & \multicolumn{2}{c|}{Text modality fine-tune} & \multicolumn{3}{c|}{Categorical} & \multicolumn{4}{c}{CCC}          \\
                       \hline
\rowcolor[HTML]{F2F2F2} 
Method                 & Stage 1     & Stage 2     & Stage 1    & Stage 2  & F1-Macro & F1-Micro & Acc.  & Val.   & Aro.   & Dom.   & Avg   \\
\hline
Baseline \cite{Naini_2025}               & \cmark                    & N/A                & N/A                     & N/A        & 0.330     & 0.475     & 0.475 & 0.653 & 0.670 & \textbf{0.604}  & 0.642 \\
Cross Attention        & \xmark                    & \cmark                  & \xmark                 & \xmark & 0.340     & 0.551     & 0.550 & 0.619 & 0.620 & 0.505 & 0.581 \\
Concat                 & \cmark         & \xmark                  & \xmark                 & \xmark            & 0.345    & 0.546    & 0.546 & 0.626 & 0.634 & 0.556 & 0.605 \\
Concat (Mish) & \cmark         & \xmark                  & \xmark                 & \xmark                     & \textbf{0.355}    & \textbf{0.605}     & \textbf{0.605} & 0.668 & \textbf{0.683} & 0.597 & \textbf{0.649} \\
Concat (Mish)          & \cmark         & \xmark                  & \cmark      & \xmark              & 0.286    & 0.493     & 0.493 & \textbf{0.682} & 0.674 & 0.588 & 0.648\\
\bottomrule
\end{tabular}
\caption{Emotional attributes and categorical results of distinct fusion strategies.}
\vspace{-3mm}
\label{tab:vad_result}
\vspace{-3mm}

\end{table*}

\section{Methodology}

This study proposes a two-stage dual-modality framework for emotion recognition, where speech and text-based features are independently fine-tuned in the first stage and subsequently fused in the second stage. The complete framework is illustrated in Figure \ref{fig:flowchart}.

\subsection{Feature Engineering}

In order to capture both the acoustic and semantic elements of emotion, we utilized both speech and text foundation features. These components are extracted through specific feature extractors designed for each modality.
\vspace{-2mm}
\subsubsection{Speech Foundation Extractor}
\vspace{-1mm}
Through the speech modality, acoustic features are extracted using WavLM \cite{chen2022wavlm}, a self-supervised learning model designed for raw audio processing. WavLM provides robust representations of prosodic and non-verbal characteristics, and has demonstrated strong performance in SER tasks within the \emph{Speech Processing Universal Performance Benchmark} (SUPERB) \cite{yang21c_interspeech}. Additionally, as WavLM is pre-trained on noisy speech, it is well-suited for capturing robust acoustic representations that generalize effectively to real-world scenarios. 

\vspace{-2mm}
\subsubsection{Text Foundation Extractor}
\vspace{-1mm}
To extract semantic features from the transcribed speech, we use RoBERTa \cite{misra2019mish}, a transformer-based model built on BERT architecture. RoBERTa excels at capturing deep contextual relationships in text, enabling it to understand word dependencies, sentence structure, and emotional nuances present in the speech content. Furthermore, it has also shown promising results in both single-modality and multi-modality SER tasks \cite{adoma2020comparative, khurana2022robinnet}.


\vspace{-2mm}
\subsection{Training Strategy}
\subsubsection{Two-Stage Scheme}
The training strategy is divided into two distinct stages to optimize the performance of both the individual modality extractors and their integration.


In the first stage, audio-based and text-based foundation extractors are fine-tuned independently for emotion recognition, enabling each modality to learn domain-specific features from its respective input. To maintain a consistent feature size, Attentive Statistical Pooling \cite{okabe18_interspeech} is applied in the speech modality, while mean pooling is used in the text modality for training the classification head. This stage establishes a robust foundation for the subsequent fusion of both modalities.


In the second stage, features for both the audio and text modalities are fused for training on the emotion recognition task. The fusion layer combines the pre-trained features from stage 1, allowing the model to learn the optimal integration of these modalities for downstream emotion recognition. The focus in this stage is to effectively combine the rich, complementary information provided by both modalities to improve the overall accuracy and robustness of emotion recognition.
\vspace{-2mm}
\subsubsection{Fusion Strategies}
To integrate the dual-modality features, the model includes a fusion layer followed by two fully connected layers: one that preserves input dimensionality and another for the downstream emotion recognition task. The fusion layer uses a simple concatenation method to combine both feature sets into a single vector. We adopt Mish activation \cite{misra2019mish}, which has recently gained attention in speech synthesis applications \cite{casanova21b_interspeech, min2021meta}, in place of the commonly used ReLU in emotion recognition tasks. Unlike ReLU, Mish is a smooth, continuously differentiable activation function that improves gradient flow, prevents neuron saturation, and enhances feature representation. 

\vspace{-2mm}
\subsubsection{Sample Balancing Control}
In large in-the-wild datasets, class imbalance often poses a significant challenge for effective model training. To address this issue, we implement \textit{weighted cross-entropy} (WCE) loss and compare it to balanced sampling and focal loss \cite{Lin_2017_ICCV} strategies. WCE loss assigns higher weights to minority classes, ensuring that their contributions to the overall loss function are amplified during training. The balanced sampling strategy mitigates imbalance by over-sampling minority classes, ensuring that each batch contains an equal number of samples per class. In contrast, focal loss prioritizes underrepresented classes by down-weighting the loss of well-classified samples, placing greater emphasis on harder-to-classify instances. This approach helps the model focus on minority classes, improving recognition accuracy for less frequent emotions and enhancing overall model robustness.

\vspace{-4mm}
\section{Experimental Settings}
\vspace{-1mm}
\subsection{Data preparation}

We conduct all experiments using the Speech Emotion Recognition in Naturalistic Conditions Challenge dataset \cite{Naini_2025}, which includes two task settings: categorical emotions and emotional attributes. In both tasks, the training, development, and testing sets are predefined by the organizers. The leaderboard evaluation is conducted on the unleashed \textit{Test3} set, while in this study, we use \textit{Test1} to evaluate our experimental results.


For categorical emotions, the dataset includes eight emotion classes: anger, contempt, disgust, fear, happiness, neutral, sadness, and surprise. It consists of 66,992, 25,258, and 24,117 in training, development, and \textit{Test1} samples, respectively. For emotional attributes, the dataset provides labels for arousal, valence, and dominance, with 84,260 training samples, 31,961 development samples, and 30,647 \textit{Test1} samples. 



To ensure a fair comparison, we strictly follow the same dataset partitioning as the baseline model. Throughout this manuscript, we use 'A', 'C', 'D', 'F', 'H', 'N', 'S', and 'U' to represent anger, contempt, disgust, fear, happiness, neutral, sadness, and surprise, respectively.

\subsection{Experimental Settings}

We carry out the two-stage training strategy on both multi-modal SER tasks (categorical emotions and emotional attributes). In stage 1, we fine-tuned RoBERTa and WavLM separately on the MSP-Podcast dataset using a batch size of 32 and a learning rate of $10^{-5}$ for 20 epochs. In stage 2, we froze both the text and acoustic feature extractors trained in the stage 1, fused their embeddings, and trained the classification head for 5 epochs with a learning rate of $5\times10^{-6}$ and a batch size of 32\footnote{The code is available at: \url{https://github.com/RogerTzeng/biic_IS2025_SER_Challenge}}.



\begin{table}[t]
\centering
\setlength{\tabcolsep}{1.pt}
\begin{tabular}{l|c|c|ccc}
\rowcolor[HTML]{F2F2F2} 
\toprule
Speech & Fine-tune & Schema   & F1-Macro & F1-Micro & Acc.  \\
\hline
WavLM  & \cmark         & WCE             & \textbf{0.338}    & 0.487    & 0.487 \\
WavLM  & \cmark         & Balanced Sample & 0.321    & 0.445    & 0.445 \\
WavLM  & \cmark         & Focal Loss      & \textbf{0.338}    & \textbf{0.592}    & \textbf{0.592}\\
\bottomrule
\end{tabular}
\caption{Categorical results of distinct balancing schema while using a fine-tuned WavLM and a pre-trained RoBERTa.}
\label{tab:cat_loss_result}
\vspace{-6mm}
\end{table}

\section{Experimental Results and Analysis}
In this section, we revisit several key training strategies in the machine learning pipeline, illustrated in Figure~\ref{fig:Common}, within the context of large-scale, naturalistic SER in the wild. Through the following subsections, we address several critical questions and challenges encountered during our participation in the competition, offering insights from multiple perspectives.

\subsection{Are Fusion Mechanisms Consistently Effective Across Different Emotion Representations?}

By using concatenation fusion with fine-tuned speech and text foundation extractors and Mish activation, our model achieves the best valence performance. To further evaluate its effectiveness in categorical emotion recognition, we compare it against alternative fusion strategies. Table~\ref{tab:vad_result} presents the results, showing that concatenation fusion with Mish activation outperforms both the baseline model and the cross-attention mechanism, achieving the highest F1-Macro and F1-Micro scores. The model using concatenation with Mish activation, fine-tuned WavLM, and pre-trained RoBERTa yields the best average attribute CCC and the highest F1-Macro. However, applying Mish activation with fine-tuned RoBERTa results in suboptimal performance in both categorical and dimensional SER tasks. We hypothesize that this degradation stems from overfitting during fine-tuning, where RoBERTa learns features highly correlated with valence, limiting its generalization to other emotional attributes and inflating valence scores in dimensional predictions. Nevertheless, these findings reinforce that simple concatenation with Mish activation is not only effective but also more robust in handling the complexities of naturalistic SER-in-the-wild scenarios.

\subsection{How Balancing Strategies Affect Speech Emotion Recognition?}



Given the highly imbalanced distribution in speech emotion classification, we evaluate the effectiveness of three balancing strategies under naturalistic SER conditions: WCE, balanced sampling, and focal loss. For balanced sampling, we implement a custom dataloader that ensures each emotion category is sampled in equal proportions within each training batch.

Table~\ref{tab:cat_loss_result} shows that focal loss achieves the best overall performance, improving F1-Micro by 10.2\% compared to weighted cross-entropy loss. This improvement aligns with expectations, as the gamma parameter in focal loss acts as a modulating factor, reducing the influence of well-classified samples and emphasizing harder-to-classify and minority categories. Unlike weighted sampling, which implicitly up-samples minority classes but lacks variability, focal loss dynamically adjusts the learning focus, ensuring better feature discrimination.

Furthermore, our results indicate that focal loss preserves accuracy for the majority classes while significantly enhancing recognition of minority emotions. This balance makes it particularly effective for highly imbalanced SER tasks, especially in in-the-wild settings where class distributions are inherently skewed. These findings suggest that focal loss provides a more adaptive and robust solution for emotion recognition under real-world conditions.


\begin{table}[t]
\centering
\setlength{\tabcolsep}{3pt}
\begin{tabular}{l|ccc|ccc}
\rowcolor[HTML]{F2F2F2} 
\toprule
Method     & F1-Macro & F1-Micro & Acc. & Val. & Aro. & Dom. \\
\hline
RoBERTa     & \textbf{0.260}     & \textbf{0.402}     & \textbf{0.402} & \textbf{0.514}     & \textbf{0.324}     & \textbf{0.308}\\
Llama-3.2   & 0.161     & 0.320     & 0.310 & 0.068     & 0.016     & 0.202\\
\bottomrule
\end{tabular}
\caption{Categorical emotion and emotional attribute results of inferring frozen text modality foundation models. }
\label{tab:cat_reg_text_result}
\vspace{-6mm}
\end{table}

\subsection{Are Pre-trained LLMs Effective for Speech Emotion Recognition in the Text Modality?}




In this experiment, we compare the effectiveness of the LLMs and the SSL for SER in the text modality. Specifically, we compare Llama-3.2 \cite{dubey2024llama} with RoBERTa-large to assess their performance in both categorical emotion classification and emotional attribute regression, as shown in Table~\ref{tab:cat_reg_text_result}.

For RoBERTa-large, we extract embeddings from the pre-trained model and pass them through two fully connected layers for classification and regression. In contrast, we use Llama-3.2 in a zero-shot setting with a carefully designed prompt, as elaborated below, for both tasks.

\begin{itemize}
    \item Categorical emotion prompts: 
    \vspace{-1mm}
    \begin{tcolorbox}[colback=gray!10,colframe=black,width=\linewidth]
    \ttfamily \scriptsize
    Predict the emotion label of the following sentence from a podcast recording. Allowed predicted emotions: ['Anger', 'Contempt', 'Disgust', 'Fear', 'Happiness', 'Neutral', 'Sadness', 'Surprise']\\
    Transcription: $<transcript>$\\
    Just predict the answer without explanation.\\
    Answer: 
    \end{tcolorbox}
    \item Emotional attributes prompts:
    \vspace{-1mm}
    \begin{tcolorbox}[colback=gray!10,colframe=black,width=\linewidth]
    \ttfamily \scriptsize
    Predict the emotional attribute label (valence, arousal, dominance) of the following sentence from a podcast recording.\\Allowed predicted ranges are from 1 to 7.\\ Transcription:$<transcript>$\\ Just predict the answer in the format of [arousal, valence, dominance], e.g., [1.0, 2.3, 4.7], without explanation.\\ Answer:
    \end{tcolorbox}

\end{itemize}


Our results indicate that RoBERTa-large consistently outperforms Llama-3.2, particularly in regression tasks, highlighting RoBERTa’s robustness and its ability to learn precise distinctions when fine-tuned for downstream tasks. While Llama-3.2 achieves reasonable performance in classification, its regression results are significantly worse. We hypothesize that while LLMs excel in categorical emotion classification, they struggle with predicting precise emotional attributes, as this requires a deeper understanding of nuanced emotional variations. Without fine-tuning or in-context demonstrations, they have difficulty inferring exact values within a restricted range. These findings suggest that for SER in the text modality, RoBERTa-large remains a more suitable choice than direct inference using a generative LLM, particularly for fine-grained regression tasks.


\begin{figure}[t]
  \centering
  \includegraphics[width=\linewidth]{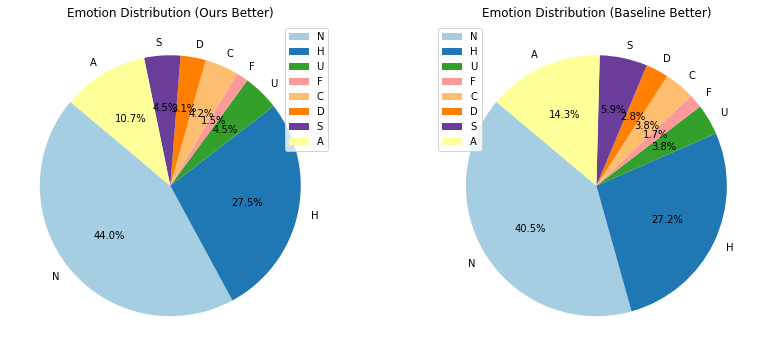}
  \caption{Pie chart of emotion categories}
  \vspace{-3mm}
  \label{fig:pie_chart}
\end{figure}


\begin{table}[t]
\centering
\begin{tabular}{cccc}
\toprule
         & {[}1, 3) & {[}3, 5) & {[}5, 7) \\
         \hline
our best & 0.0553   & \textbf{0.4492}   & 0.1956   \\
\hline
baseline & 0.0672   & 0.3947   & 0.1975  \\
\bottomrule
\end{tabular}
\caption{valence CCC among certain ranges}
\label{tab:split_ccc}
\vspace{-3mm}
\end{table}

\subsection{Quantitative Analysis}
To further understand the improvements in valence prediction comprehensively, we first carry out an analysis of the statistical distribution of predictions on the Test1 dataset with the mean prediction and \emph{standard deviation} (mean $\pm$ std). The baseline model yields a mean prediction of 3.78$\pm$1.0, while our best approach produces a mean of 3.83$\pm$0.92. Compared to the ground truth (3.89$\pm$1.0), our model demonstrates a closer mean prediction and lower variance, suggesting a better fit to the data and improved stability.

Furthermore, we compare samples where our best model outperforms the baseline, categorized by discrete emotions. Figure~\ref{fig:pie_chart} illustrates the emotion distribution among these samples. Since valence prediction is a regression task, we compute the MSE loss for both models against the ground truth and filter samples based on lower MSE values (indicating better predictions). Our analysis reveals that our best approach achieves notable improvements in recognizing surprise (4.5\% vs. 3.8\%) and contempt (4.2\% vs. 3.8\%), two emotions that are typically more challenging to classify, apart from fear and disgust. This suggests that our model effectively leverages text-based features to enhance valence prediction.


Lastly, to quantitatively validate our model’s superiority, we segment the samples into three equal-range bins based on the ground truth: [1, 3), [3, 5), and [5, 7). The CCC scores of each partition are reported in Table~\ref{tab:split_ccc}. Notably, our model outperforms the baseline most significantly in the [3, 5) range (0.449 vs. 0.395), suggesting that it provides more accurate predictions in this middle interval, thereby ensuring a smoother distribution across the entire valence spectrum.

\begin{table}[h]
\centering
\begin{tabular}{l|c}
\toprule
Team Name & Valence \\
\hline
biic      & $\textbf{0.6953}^{\star}$  \\
Voinosis  & 0.6941  \\
SRPOL     & 0.6932  \\
SRPOL     & 0.6906  \\
biic      & 0.6880  \\
\bottomrule
\end{tabular}
\caption{Top-5 rank in emotional attributes (valence) among the \textit{Test3} testing set.}
\label{tab:final_rank_result}
\vspace{-10mm}
\end{table}

\vspace{-2mm}
\section{Conclusions}


Through our experiments and dual-modality fusion architecture, we achieved the best performance on \textit{Test3}, obtaining a valence CCC of 0.6953 (Table~\ref{tab:final_rank_result}). Our findings suggest that focal loss and activation functions play a crucial role, particularly in valence regression, even without employing excessively deep or complex models. Additionally, we observe that the concatenation fusion strategy with Mish activation consistently enhances performance across both categorical emotion classification and emotional attribute regression.
Furthermore, while LLMs excel in various domains, our results indicate that without fine-tuning or in-context learning, their performance in SER-in-the-wild remains inferior to traditional, robust text models.

However, our model exhibits limited improvements in categorical emotion, arousal, and dominance prediction, suggesting that a more generalized SER framework remains an open challenge. Future work should focus on a deeper analysis and more comprehensive modeling approaches to improve SER performance in naturalistic, real-world conditions.
\bibliographystyle{IEEEtran}
\bibliography{mybib}

\end{document}